# Synthesis and characterization of core-shell structure silica-coated $Fe_{29.5}Ni_{70.5}$ nanoparticles

M Ammar[1], F Mazaleyrat[1,4], J P Bonnet[2], P Audebert[2], A Brosseau[2], G Wang[3] and Y Champion[3]

[1]SATIE, ENS Cachan, CNRS, UniverSud, 61 av President Wilson, F-94230 CACHAN, France.

[2]PPSM, ENS Cachan, CNRS, UniverSud, 61 av President Wilson, F-94230 CACHAN, France.

[3]CECM CNRS-UPR 2801, 15 rue Georges Urbain 94407 Vitry-sur-Seine, France

[4]IUFM de Créteil, rue Jean Macé, F-94861 Bonneuil-sur-Marne, France

E-Mail : ammar@satie.ens-cachan.fr

**Abstract**

In view of potential applications of magnetic particles in biomedicine and electromagnetic devices, we made use of the classical Stöber method – base-catalysed hydrolysis and condensation of tetraethoxysilane (TEOS) – to encapsulate FeNi nanoparticles within a silica shell. An original stirring system under high power ultrasounds made possible to disperse the otherwise agglomerated particles. Sonication guaranteed particles to remain dispersed during the Stöber synthesis and also improved the efficiency of the method. The coated particles are characterized by electron microscopy (TEM) and spectroscopy (EDX) showing a core-shell structure with a uniform layer of silica. Silica-coating does not affect the core magnetic properties. Indeed, all samples are ferromagnetic at 77 K and room temperature and the Curie point remains unchanged. Only the coercive force shows an unexpected non-monotonous dependence on silica layer thickness.

## 1. Introduction

Magnetic-metal nanoparticles encapsulated in a dielectric inorganic material are considered to have practical applications in electromagnetic devices, biology and fundamental study to improve the local physical investigation of magnetic nanostructures. In the core-shell structure, the core size-dependant magnetic susceptibility at room temperature combined with the chemical stability of the silica coatings suggests that the resulting nanocomposite may be a good candidate for biomedical applications, such as magnetic separation, drug targeting, image contrast in magnetic resonance imaging and hyperthermia therapy [1,2,3]. Magnetic fluids dedicated for clinical applications are typically colloidal suspensions of iron, magnetite, iron-nickel and cobalt nanoparticles coated with biocompatible surfactants [4]. Actually, there are two fundamental criteria to prevent the catalysis of damaging reactions within cells, the reduction of the toxicity of the vector conveying the solution due to its oxidative alteration and its chemical time stability. Accordingly, the silica coating of magnetic nanoparticles is one of promising tool to ensure this specific biocompatibility and leads to low toxicity material.

Magnetic-dielectric nanocomposites have also attracted sustained interest over one century owing to their unusual combined magnetic and electric properties. In fact, due to their metallic nature, eddy currents limit application of magnetic nanoparticles at high frequency. The coating by an insulating shell on the surface of soft magnetic nanoparticle cores such as FeNi confers to the material a high permeability independent of the frequency even in GHz range [5]. Such materials are typically suited for applications in telecommunication [6]. On the other hand, the ability to control magnetic interactions is an important consequence of the coating of magnetic particles, which has been explored in details by several authors for particles in solution [7] and close-packed thin films [8]. Coating thickness controls both insulation of nanoparticles and interparticle distance and, therefore, the interparticle interactions [9]. This provided substantially reliable results to study magnetic nanostructure of nanoparticles using electron holography [10].

Several synthetic routes for producing magnetic nanoparticles have been explored during the last decade including chemical vapor condensation (CPVD), powder pyrolysis and sonochemical synthesis [11,12,13]. However, nanoparticles synthesized by these methods frequently display a relatively poor cristallinity or polydispersity in their shape or/and size, which affects their magnetic properties. Evaporation–cryo-condensation process has been developed to overcome some of these problems. In the present work, cryogenic melting has been used to produce $Fe_{29.5}Ni_{70.5}$ nanoparticles and consequently to guarantee more cristallinity and a better stability in the elemental chemical composition.  Additionally, several approaches, such as the sol-gel process [14], co-precipitation [15,16,17], metal-dielectric co-sputtering deposition [18] or metal ion deposition, have been used to prepare magnetic-insulator nanocomposites. Our present approach is to start from metallic nanoparticles and to coat them with an inorganic-dielectric polymer in order to control the morphology of the shell. In this paper, a modified Stöber approach has been used to encapsulate in silica the as-prepared metallic FeNi particles. In fact, we have introduced high-power sonochemistry not only in the dispersion step, but also during the synthesis to improve the effectiveness of the classical Stöber method [19,20].

## 2. Experimental details

*Synthesis of $Fe_{29.5}Ni_{70.5}$ nanoparticles*

FeNi nanoparticles with well-defined morphology and homogeneous chemical composition were synthesised using the cryogenic melting technique. This method consists in sliding down a feeding bar of metal ($Fe_{29.5}Ni_{70.5}$) into a Radio Frequency (RF) reactor. A drop of molten metal forms at the edge and falls onto the inductors where it is levitated to complete transformation into nanocrystalline powders. In order to have sufficient vapour pressure, the metal must be heated up by several hundred degrees over its melting temperature (over 2000°C for Fe, Ni). The size of particle depends critically on the metal vapour pressure. The gas produced from the cryogenic liquid carry the particles into a canvas filter. Technical details are reported in [21]. The as-obtained iron-nickel nanopowders are composed of spherical particles with an average diameter of about 55 nm (deduced from microscopy, standard deviation 20 nm). From Electron Energy Loss Spectroscopy (EELS) the chemical composition is homogeneous from one particle to another as well as inside the nanoparticles. The fraction of iron x = 0.295 is of particular interest since large amounts can be produced with no deviation in chemical composition [22]. Because metallic nanoparticles are pyrophoric in air, they are collected in hexane where an oxide layer of approximately 2 nm forms, making possible their manipulation without risk. The magnetization of as-prepared iron-nickel particles (75 $Am^2/kg$) is 20% lower compared to the bulk alloy magnetization, which confirms the non-magnetic nature of the oxide layer observed from electron microscopy and analysed using XPS (X-Ray Photoemission Spectroscopy). Essentially Nickel Hydroxides Ni(OOH) and Ni(OH$_2$), iron oxide $Fe_2O_3$ and FeO were detected [22].

*Synthesis of Silica-coated $Fe_{29.5}Ni_{70.5}$ nanoparticles*

The silica shell onto FeNi core was synthesized according to the Stöber method [23] (sol-gel reaction) without any silane coupling agent (like 3-aminopropyltrimethoxysilane which is sometimes used for noble metals nanoparticules silica coating [24]). Indeed, oxide shell covering FeNi nanoparticles is expected to enhance the $SiO_2$ shell binding.

Ethanol (95-96% synthesis grade) and Ammonia solution (28% analytical grade) were purchased from SDS/CARLO ERBA, tetraethylorthosilicate (TEOS) ≥ 98% (GC) from FLUKA. All reagents were used as received without further purification.

Ultrasonic dispersion was carried out with a Bandelin 200W (variable from 10 to 100%) ultrasonic processor (Sonopuls HD 2200) fitted out with a horn of 13×3 mm. All experiments were made in glass flask equipped with a cooling jacket to keep the mixture temperature constant.

Typically, 80 mg of raw $Fe_{29.5}Ni_{70.5}$ nanoparticles were first sonicated in 50 ml of ethanol during 90 minutes under a controlled ultrasonic power of 3 W/cm$^3$. Then, various volumes of TEOS and ammonia 28% ($NH_4OH$) were successively introduced into the suspension and the mixture was again sonicated for 90 minutes under a power of 0.5 W/cm$^3$ to complete the sol-gel reaction (Figure 1). Finally the suspensions are centrifuged at 3000 rpm for 10 minutes, the solvent is discarded, and the FeNi nanoparticles are ultrasonically redispersed in 50 ml of ethanol. This purification process (centrifugation/dispersion under sonication) was repeated three times. The particles were then transferred into ethanol to avoid any further growth or chemical modification of the silica layer. Subsequently, an amount of the coated nanoparticles were dried under reduced pressure and moderate temperature to remove remaining solvent and to prepare samples for the physical characterization. Four samples with different reagent concentrations have been produced (see table 1).

*Characterization and techniques*

Thermal degradation analyses were made with a Perkin-Elmer Pyris 6-TGA instrument using standard ceramic crucibles and sample mass of 1-29mg. The samples were heated at a rate of 10 °C min$^{-1}$ from room temperature to 1000 °C in an air flow of 10 ml min$^{-1}$ or an argon flow of 80 ml min$^{-1}$. The analyser was coupled to a permanent magnet producing a gradient field in the crucible to measure the Curie temperature ($T_c$). These measurements are conducted under argon flow to avoid adventitious oxidation of the nanoparticles. Fourier Transform Infra-Red spectra (FTIR) were recorded with a Thermoelectron Corporation NEXUS spectrometer equipped with an attenuated total reflectance probe (ATR) covering the wavenumber range 4000-700 cm$^{-1}$. The morphology and size of the particles were analysed by conventional and high resolution electron microscopy (HRTEM) using a TECNAI F20 microscope (operating at 200 kV with a point-to-point resolution of 0.24 nm), on the powders deposited onto a microscopy grid coated with an amorphous carbon film. Powders were also characterized by Electron Energy Loss Spectra (EELS) in a Gatan Image Filter (GIF 2000) spectrometer coupled to the TECNAI F20. Fitting and integration windows of 30 eV were used for all the chemical maps and the spectra were obtained with an energy resolution of 1.2 eV. Elemental chemical analysis of the nanocomposites was also performed using Energy Dispersive X-rays (EDX) attached to the same system. The quasi-static hysteresis loops with an applied magnetic field of –300 kA/m < H < 300 kA/m were acquired using a home-made Vibrating Sample Magnetometer (VSM) between room and liquid nitrogen temperature.

## 3. Results and discussion

Fourier Transform Infrared (FTIR) spectroscopy was used to identify the functional groups present on the surface of FeNi@SiO2 nanoparticles. Comparison of uncoated and silica-coated particles FT-IR spectra (Figure 2) shows a pronounced change detected in the 1300-700 cm$^{-1}$ region, which clearly indicates the presence of the silica coating. The peaks at 970 and 1070 cm$^{-1}$ correspond to the characteristic Si-O-Si bond, typically attributed to the Si-O$^-$ symmetric stretching and Si-O-Si asymmetric stretching respectively, in agreement with [25]. Analysis of bonding configurations from FTIR data suggest also the existence of Si-O-C or/and ≡Si-O-Si≡ functions (bands under 1000 cm$^{-1}$). Nevertheless, the spectra are obviously dominated by the Si-O-Si bonding vibrations, for all coated samples. The presence of this type of strained bond is a clear evidence of the mechanical stress in the silica sheath, which in turn may strain the FeNi nanoparticles.

The chemical composition was examined using Energy–Dispersive X-ray (EDX) spectroscopy, which shows a Fe$_{29.5}$Ni$_{70.5}$ core coated by silica shell (Figure 3). The copper lines in this figure are due to the copper grid used as TEM sample holder. An atomic ratio of Si/O = 1/0.6 was obtained on the core-shell structure, indicating that the off-stoichiometric silica shell is silicon-rich in nature. The structural evolution study suggests that the silica layer grows without affecting the integrity of the FeNi core. Indeed, the spectra do not reveal other elements except those present initially in the FeNi core, the oxide layer, the silica shell and the copper grid.

Figure 4 shows TEM pictures of FeNi@SiO$_2$ particles synthesized using various TEOS volumes. Observation of figure 4 images (a), (b), (c) and (d) clearly shows the shell thickness dependence on TEOS concentration (see also table 1). Additionally Energy Filtered in scanning TEM mode, which one can see an illustration on the inset (f) of figure 4, comes to support the elementary chemical nature of the silica-layer surrounding the nanoparticles. In fact, the image exhibits a chemical cartography obtained from EELS and undeniably shows the formation of silica uniformly on the FeNi core.

The properties of oxidation-resistance of the FeNi@SiO$_2$ composite were tested by TGA. Figure 5 shows the typical curves of thermal analysis of metallic materials [26]. Correspondingly, the weight increment of the coated particles (sample 5) caused by FeNi oxidation decreased from 28% to 5% relative to that of the uncoated FeNi particles (sample 1). It is clear that a thicker shell of silica can protect the nickel-iron from oxidation more efficiently. For instance the oxidation of the FeNi core of FeNi@SiO$_2$ composites (sample 4) proceeds at ~430 °C which is 250 °C higher than for as-prepared FeNi nanoparticles. The weight loss, observed for coated samples starting from RT, is attributed to the surface dehydration of the silica monolayer and the loss of others organic compounds which are volatile in this range of temperature [27].

For many applications of core-shell particles, such as electromagnetic devices [28], it is of essential importance to control precisely the thickness of the shell. In the system under consideration, the simplest approach to vary shell thickness is to use different amounts of TEOS. Consequently, we investigate the effect of adding various amounts of TEOS in a single step. Figure 4 (e), which features a typical high resolution image (HRTEM) for sample 4, reveals a core-shell structure with an uniform amorphous silica coating (thickness 15 nm). For comparison, the thickness of silica-shell is deduced from magnetic characterization. In fact, the volume of $SiO_2$ can be estimated using

$$V_{SiO_2} = \frac{m_{SiO_2}}{\rho_{SiO_2}} = \frac{m_{Coated}}{\rho_{SiO_2}} \left( 1 - \frac{M_{s(Coated)}}{M_{s(Free)}} \right)$$

where $V$ and $m$ are for volume and weight, respectively, $\rho_{SiO2}$ is the silica density estimated experimentally (2270 kg/m$^3$) and $M_s$(Am$^2$/kg) is the specific magnetic moment at saturation; "coated" indicates the coated sample, and "free" corresponds to the raw FeNi powder. Assuming that the nanoparticles are monodisperse and 55 nm in diameter, the theoretical thickness $t_{MAG}$ required to increase the radius R of the seed particle to a final radius $R+t_{MAG}$ is given by [29]

$$t_{MAG} = R\left(\sqrt[3]{\left(\frac{V_{SiO_2}\rho_{FeNi}}{m_{FeNi}}+1\right)}-1\right)$$

where $m_{FeNi} = m_{Coated} - m_{SiO2}$ is the weight of the effective magnetic component in the nanocomposite. The bulk $Fe_{29.5}Ni_{70.5}$ density was used (8450 kgm$^{-3}$ [30]).

Figure 6 shows the dependence of the thickness of silica shell, deduced from TEM analysis and magnetic measurements, on TEOS volume. Interestingly, the two data are consistent with a quantitative silica formation on the nanoparticles for thin silica layer up to 20 nm. Above this limit $t_{MAG}$ presents a discrepancy compared to $t_{TEM}$ for thicker silica layer (beyond 20 nm). This could be explained by the presence of free silica nanoparticles synthesized when a large amount of TEOS is added. After centrifugation, the calculated volumic amount of silica coating the nanoparticles is underrated and therefore, the deduced silica-shell thickness is erroneous.

4. **Magnetic properties**

The TGA recordings under constant magnetic field are presented in figure 7. Due to the neutral atmosphere (argon flow), oxidation was inhibited. Up to 600 °C we observe a weak weight drop due to a chemical desorption from the silica shell for coated particles as reported in [27]. Comparable weight loss is observed for uncoated sample 1 due to the desorption of organic chains adsorbed in the oxidized FeNi surface during the passivation step of the nanoparticles. The TGA traces show a characteristic feature for all samples which reveals a typical ferromagnetic-to-paramagnetic transition at the Curie temperature ($T_c$). Noticeably, the nanocomposites (sample 2 to 5) exhibit a broaden transition. Obviously, this makes difficult the extraction of $T_c$ which roughly maintains a stable value of 605 °C (± 5 °C) for all samples in agreement with the literature [31].

A comparative measurement of hysterisis loops at 300 K (RT) and 77 K was performed for both uncoated and silica-coated nanoparticles using VSM as mentioned previously. Magnetization curves are reported in figure 8 and the main quantities are listed in table 2 (specific saturation magnetization, remanent magnetization and coercivity at 77 K and RT). All curves at RT saturate approximately at the same applied field than those measured at 77 K. For the same operating temperature, loops for coated samples appear to have a component whose magnetization continues to increase with increasing field up to 200 kA/m, whereas the raw FeNi powder saturates much faster than the nanocomposites. In fact the interparticle interactions are modulated by the thickness of the coating layer which isolates the particles. As a result the nanocomposite hardens magnetically and its saturation becomes difficult [6,18,32]. For all samples there is only a slight deviation regarding the saturation magnetization between 77 K and RT because RT/$T_c$ ≈ 0.3. Furthermore, the coating quality is examined in saturation magnetization versus silica-layer thickness plots as an inset in figure 8 (right side). It is clearly seen that the specific magnetization decreases with increasing the thickness of silica-shell. Accordingly diamagnetic contribution of silica leads to a lower saturation magnetization than the core-free FeNi particles (table 2).

The inset, left side of figure 8, illustrates the coercive field versus the thickness of silica-shell at 77 K and RT. Noticeably, the temperature dependence of coercivity indicates a slight increase for all samples when temperature decreases which is consistent with an increase of anisotropy regardless of its origin. On the one hand, for randomly oriented nanoparticles with cubic anisotropy, the coercive field should be $H_c \approx 0.64 K_1/J_s$ [33]. If we consider the bulk $Fe_{30}Ni_{70}$ magnetocristalline anisotropy $K_1 \approx 700$ J/m$^3$ [34] and the measured saturation magnetization $J_s = 0.8$ T, we find $H_c \approx 560$ A/m which is in disagreement with the experimental coercivity. On the other hand, the morphology and size effect are believed to be the reason of high coercivity observed for all samples (22 kA/m < $H_c$ < 32 kA/m, see table 2). According to the pioneering work of Néel [35], for soft magnetic nanoparticles a dissymmetry of some atomic layers is sufficient in order to make the contribution of demagnetizing field becoming dominant and to lead to an enhancement of the coercivity. Shape anisotropy effect is due to the asphericity of the nanoparticles below a critical size. The coercive field in an elongated spheroidal single-domain particle is given by $H_c = 2K_s/J_s$ [36] where $K_s = \dfrac{(N_b - N_a)J_s^2}{2\mu_0}$ is the shape anisotropy. $N_b$ and $N_a$ represent the demagnetizing coefficients along the two axes of an ellipsoid of revolution [37]. For an asphericity 0.86<$\gamma$<1.14, $N_a = \dfrac{1}{3}\left(\dfrac{9}{5} - \dfrac{4}{5}\gamma\right)$ and since $N_a + 2N_b = 1$ we find $K_s = \dfrac{(\gamma-1)J_s^2}{5\mu_0}$ ($\gamma$>1).

Consequently, we deduce

$$H_c = \frac{2(\gamma-1)J_s}{5\mu_0}$$

If we assume a coercive field $H_c$ = 33 kA/m, we find an asphericity of 13%. The result is consistent with TEM observations (figure 4) where 0.80<$\gamma$<1.2 was found. The coercive field dependence on silica layer thickness shows a non-monotonous evolution. For thin silica-layer a sensitive drop in coercivity is observed followed by an increase before recovering the initial value. This is probably due to a competition between dipole-dipole interaction and magneto-elastic anisotropy. In the one hand, dipolar interactions are reduced as the distance between magnetic cores is increasing. In the other hand, it has been shown by FTIR the existence of stress in the silica shell. As the thickness of the shell increases, the stress experienced at the surface of FeNi nanoparticles is enhanced yielding an increasing magneto-elastic anisotropy. These two contributions balanced for a thickness of ~15 nm. Classically, for ultrafine nanoparticles (~10 nm or less) dispersed in non-magnetic material, anhysteretic loops are expected because of the superparamagnetic behavior of the nanoparticles, as already reported [38,39]. For the FeNi nanoparticles described in this paper, the shape anisotropy dominates the magnetocristalline anisotropy ($K_s$ = 20 kJ/m$^3$ >> $K_1$) so the critical size for which the superparamagnetism is observed at room temperature is given by

$$D_{sp} = \sqrt[3]{\frac{150 k_B T}{\pi K_s}} \approx 27 \; nm$$

compared with $\sqrt[3]{\frac{150 k_B T}{\pi K_1}} \approx 72\ nm$ [40], where $T$ is the measuring temperature and $k_B$ the Boltzmann constant. Considering the mean size of the nanoparticles (55 nm), this is in line with the hysteresis observed for all samples (figure 8). Because part of the particles is smaller than 27 nm and because some are nearly spheroidal ($\gamma<1.05$) a superparamagnetic contribution is not excluded. Another interesting feature is the remarkable stability of the squareness ratio regardless of temperature and coating (see table 2). The $M_r/M_s$ ratio is noticeably lower than 0.5 predicted for single domain particles according to Néel and Stoner [41,42]. Actually, this low value is typical of vortex-like magnetic structure composed of an out of the plane uniformly magnetized core surrounded by a crown of curling spins [10]. Alternatively to a coherent rotation, the magnetization process consists initially into an irreversible switch of the vortex core followed by a screw-like rotation of the external curling spins [43].

## 5. Conclusions and perspectives

The preparation of silica-coated FeNi particles was successfully achieved by a combination of two original synthetic procedures, a cryogenic evaporation of master alloy $Fe_{29.5}Ni_{70.5}$ to obtain nanoparticles with well-defined size and composition, and subsequently a modified classical Stöber method which permits to encapsulate the latter within a silica shell. The coating can be accomplished through a direct, simple, one-step procedure. FTIR, EDX and EELS analysis are consistent with the presence of silica in the nanocomposites synthesized. Consequently the silica-shell thickness could be conveniently controlled through the TEOS volume added to the colloidal FeNi solution. Our study allowed us to correlate the shell-silica thickness with the evolution of the magnetic properties of the final nanocomposite. The magnetic investigations demonstrate the possibility of making property-tunable magnetic nanoparticles ready for surface engineering in particular with bioactive molecules or for electromagnetic device applications aiming to enhance frequency limits. These aspects will undoubtedly require further longer-term ageing studies. In particular, the chemical stability must be ensured before any in-vivo applications are intended. Electronic holography experiments are in course to confirm the expected vortex structure of the FeNi nanoparticles.

**Acknowledgements**

This work was supported by the Institut d'Alembert IFR-CNRS-UMR 8531, 61 Av. Du President Wilson 94235 Cachan, France.

**Figure Captions**

**Figure 1.** Illustration of the silica-coated $Fe_{29.5}Ni_{70.5}$ nanoparticles stepwise synthesis protocol.

**Figure 2.** Thermal gravimetric curves (TGA) of FeNi@$SiO_2$ nanocomposites (sample 1 to 5) under air flow. Thermal nanopowder alteration obviously depends on the amount of silica in the sample.

**Figure 3.** FTIR spectra recorded from different samples FeNi@$SiO_2$ (sample 1 to 5) related to various volumes of TEOS in range of 750-2500 $cm^{-1}$. The main resonances are identified in the figure and discussed in the text in relation with the dominating Si-O-Si vibrations on solid surface.

**Figure 4.** In the inset of top, EDX spectrum of $Fe_{29.5}Ni_{70.5}$ core-free nanoparticles. In the inset of bottom, EDX of silica portion of a FeNi@$SiO_2$ nanoparticles when the beam was focused on silica edges.

**Figure 5.** (a-d) Representative transmission electron micrographs of the silica-coated FeNi nanoparticles corresponding respectively to the sample 1, 2, 4, 5 (e) HRTEM pattern for the silica@FeNi (sample 3) which shows the presence of a 15 nm thick silica layer lying at the particle surface (f) Typical EFTEM analysis using metallic silicon as the silica source (Si K-edge), displays the chemical cartography showing a silica-rich shell (sample 3).

**Figure 6.** Plot of silica-layer thickness as function of various volume of precursor TEOS, estimated from HRTEM analysis ($t_{TEM}$) and magnetic characterization ($t_{MAG}$) (see also table 1).

**Figure 7.** Thermal gravimetric curves (TGA) of FeNi@SiO$_2$ nanocomposites (sample 1 to 5) under argon flow. See table 2 for Curie temperatures assessed from curves.

**Figure 8.** Magnetic quasi-static hysterisis loops for samples with various silica-shells (sample 1 to 5). On the left, magnetization curves recorded at 300 K. The inset on the lower right corner illustrates the changes in the $M_s$ as a function of the silica-shell thickness. On the right the M–H curves recorded at 77 K. The inset on the lower right corner illustrates the changes in the coercive field as a function of the silica-shell thickness.

**Table 1.** Summary of the FeNi@SiO$_2$ synthesis, presenting the various volumes of reagents used. The Silica-layer thickness was estimated using HRTEM analysis ($t_{TEM}$) and magnetic investigation ($t_{MAG}$).

| | Silica Coating (after dispersion under ultrasounds (3 w/cm$^3$)) | | t (nm) | |
|---|---|---|---|---|
| | TEOS(μl) | NH$_4$OH(ml) | $t_{TEM}$ | $t_{MAG}$ |
| Sample 1 | 0 | 0 | 0 | 0 |
| Sample 2 | 50 | 0.35 | 3 | 4 |
| Sample 3 | 100 | 0.7 | 8 | 9 |
| Sample 4 | 200 | 1.4 | 15 | 17 |
| Sample 5 | 500 | 3.5 | 33 | 24 |

**Table 2.** Magnetic properties for uncoated FeNi (sample 1) and silica-coated FeNi (sample 2 to 5) nanoparticles. $M_s$ is the specific saturation magnetization, $M_r$ remanent magnetization, $T_c$ Curie temperature and $H_c$ coercive field.

| | $M_s$ (Am²/kg) | | $M_r/M_s$ | | $H_c$ (kA/m) | | $T_c$ (°C) |
|---|---|---|---|---|---|---|---|
| | 77 K | 300 K | 77 K | 300 K | 77 K | 300 K | |
| Sample 1 | 80 | 76 | 0.32 | 0.34 | 33.9 | 31.8 | 598 |
| Sample 2 | 72 | 65 | 0.30 | 0.30 | 28.2 | 22.1 | 608 |
| Sample 3 | 58 | 54 | 0.26 | 0.28 | 28.8 | 24.4 | 607 |
| Sample 4 | 40 | 39 | 0.27 | 0.28 | 33.8 | 31.1 | 607 |
| Sample 5 | 32 | 29 | 0.28 | 0.31 | 34.3 | 32.2 | 607 |

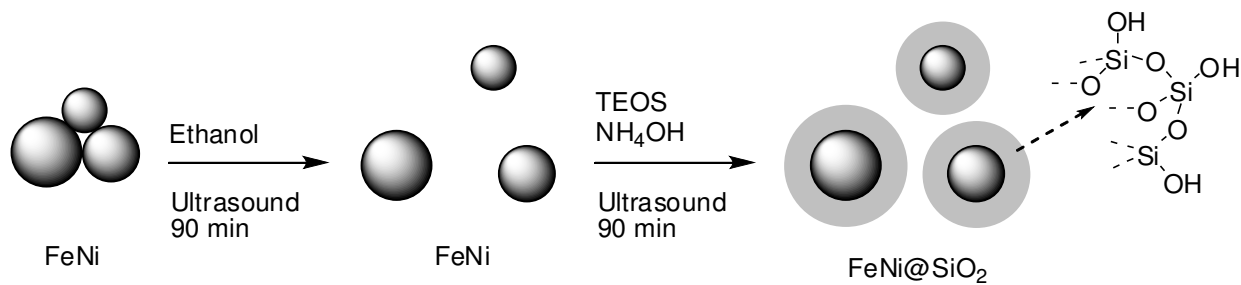

Figure 1

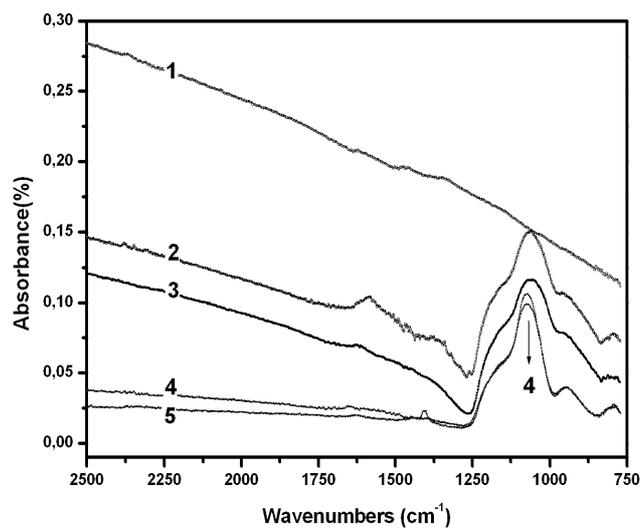

Figure 2

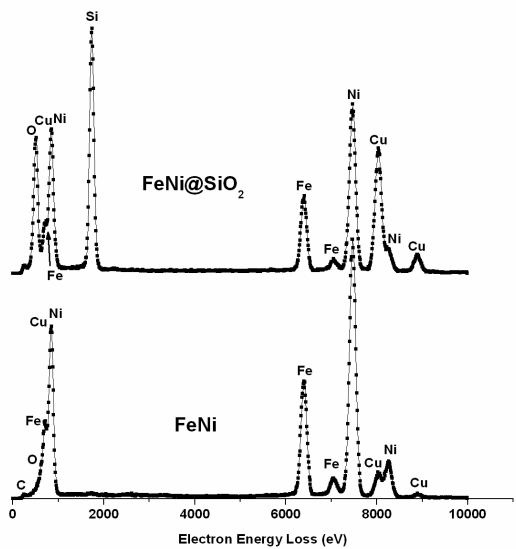

Figure 3

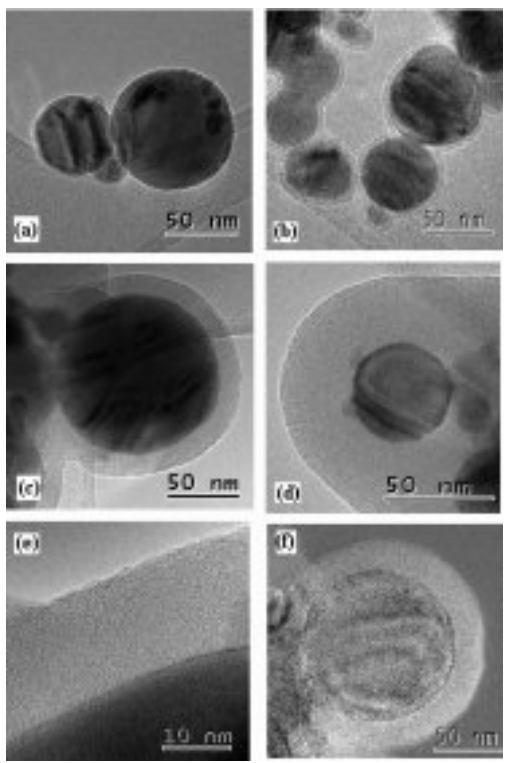

Figure 4

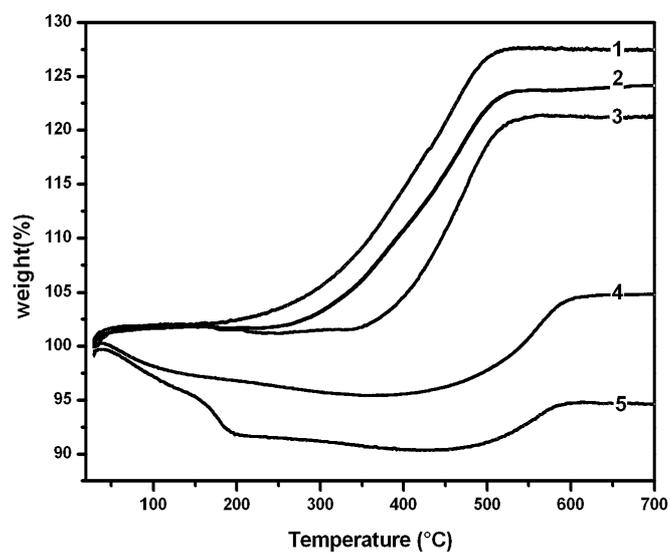

Figure 5

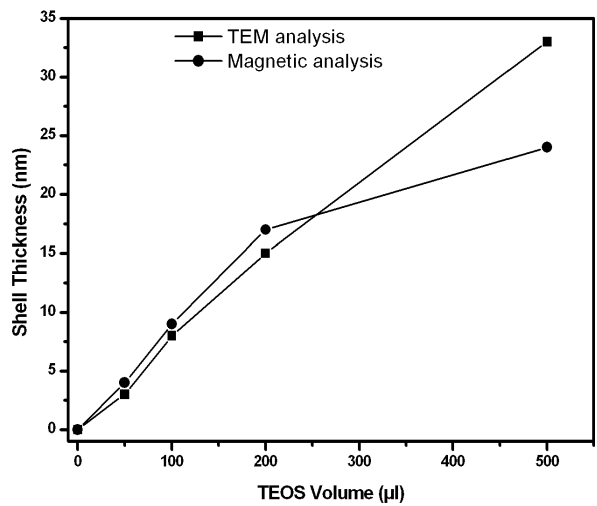

Figure 6

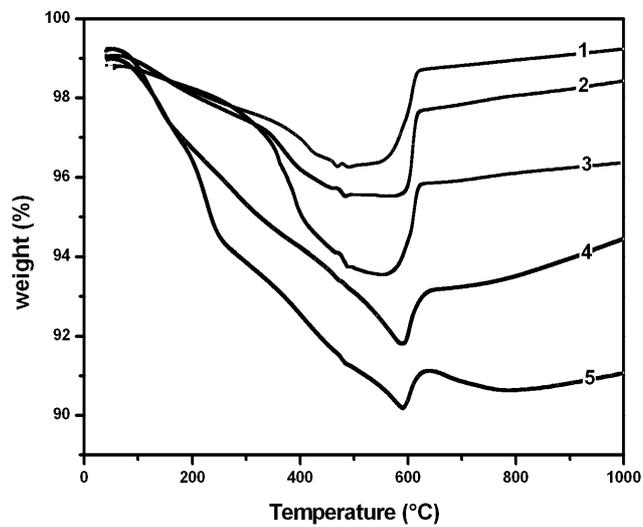

Figure 7

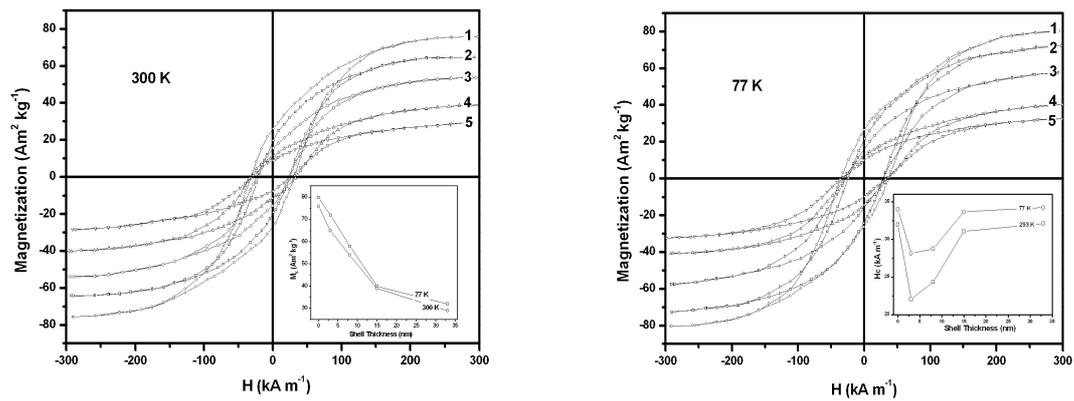

Figure 8